\documentclass[a4paper,12pt]{article}
\usepackage[top=2.5cm,bottom=3cm,width=16.5cm]{geometry}
\usepackage{amssymb}
\usepackage{color}
\usepackage{tikz,eso-pic, ulem, pifont, lscape, fancyvrb, alltt,fancybox}
\usetikzlibrary{calc,3d,arrows, plotmarks,shapes,patterns}
\title{Kinematical versus Dynamical Contractions\\ of  de Sitter Lie algebras}
\author{Joachim Nzotungicimpaye\footnote{Unit\'e de Recherche en Sciences Naturelles, ENS, Burundi}\footnote{email address: kimpayakad@gmail.com}}
\begin{document}
\maketitle
\date{}
\begin{abstract}
To better clarify the three kinds of Bacry and L\'evy-Leblond contractions \cite{jmll}:(speed-time, speed-space and space-time contraction), we first introduce kinematical parameters, namely the radius $r$ of the universe, the period $\tau$ of the universe and the speed of light $c=r\tau^{-1}$. Next we present them as static, Newtonian and flat limits through the use of the dynamical parameters, namely the   mass, $m$, the energy, $E_0$ and the compliance $C$, all depending on mass as well as length and time. To give a little physical taste for each kinematical Lie algebra, we set up equations of the change  with respect each group parameter through the use of the Poisson brackets defined by the Kirillov form.
\end{abstract}

\section{Introduction}
Group (algebra) contraction is a method which allows to construct a new group (algebra) from an old one. Contraction of Lie groups and Lie algebras started sixty six years ago with E.Inonu and E.P.Wigner \cite{inonuwigner} in $1953$, when they were trying to connect Galilean relativity and special relativity. Eight years later, in $1961$, E.Saletan \cite{saletan} provided a mathematical foundation for the Inonu-Wigner method. Since then, various papers have been produced and the method of contraction has been applied to various Lie groups and Lie algebras \cite{jmll2,sanjuan,dooley, gromov,fialowski,  mcrae,guo, ancilla1}.

The method has also been used by H.Bacry and J.M.L\'evy-Leblond \cite{jmll} to connect the de Sitter Lie algebras to all other kinematical Lie algebras through three kinds of contractions: speed-space contractions, speed-time contractions and space-time contractions. The terminology is related to the fact that Bacry and L\'evy-Leblond have, first of all, scaled the velocity-space generators, the velocity-time translation generators and the space-time translation generators by a parameter $\epsilon$ to obtain, in the limit $\epsilon\rightarrow 0$, the respective contractions that we prefer to call velocity-space contractions, velocity-time contractions and space-time contractions. The L\'evy-Leblond contraction approach has been also extended to supersymmetry \cite{campaomor2010} and kinematical superalgebras \cite{rembielinski, chao,barducci1}.

Within the corresponding eleven Lie groups, four of them, namely the Galilei group $G$ governing the Newtonian physics (Galilean relativity), the Poincar\'e group $P$ governing the Einstein physics (special relativity), the Newton-Hooke groups $NH_{\pm}$ describing  Galilean relativity in the presence of a cosmological constant and the de Sitter Lie groups $dS_{\pm}$ governing the de Sitter relativity of a space-time in expansion or oscillating universe, are well known in physics literature.

Within the remaining five ones, the Para-Poincar\'e groups $P_{\pm}$ and the Static $S$ are still unknown in physics, but the Para-Galilei group $G_\pm$ and the Carroll group $C$ are gaining more interest in recent times.

The Para-Galilei group has been identified as governing a light spring \cite{ancilla}.

The Carroll group has been associated to tachyon dynamics \cite{gibbonsthoughts,gibbonscondensates,escamilla}, to Carrollian electromagnetism \cite{duval} versus Galilean electromagnetism \cite{Lebellac} or to the dynamics of Carroll particles \cite{joakimgomis} and Carroll strings \cite{cardona}.
The anisotropic  Carroll group in two space dimensions (i.e. without rotations) has been identified as the isometry group of  gravitational plane waves \cite{Sou73,Carroll4GW}. The Carroll group has also been used recently in the study of ultra-relatistic gravity \cite{hartong} and for the generalization of  Newton-Cartan gravity \cite{bekaert1,bekaert2}.
The Carroll group has been compared to the Galilei group in the study of gravitational waves \cite{morand, DGH91}, of confined dynamical systems \cite{barducci2}, of gravity \cite{bergshoef} and of covariant hydrodynamics \cite{ciambelli}.

The purpose of this paper is, first of all, to clarify the origin of the names given to the three L\'evy-Leblond types of contraction and then improve the Levy-Leblond method further.

Firstly we note that the kinematic descriptions are associated only with lengths and times, while the dynamic descriptions are associated  with the mass as well as with lengths and time.

 In section $2$ we recall the Inonu-Wigner contraction,  while  section $3$ recalls the Bacry - L\'evy-Leblond method and uses it to establish the twelve kinematical Lie algebras as obtained by A. Ngendakumana and al.\cite{ancilla1}. With section $4$ we revisit the L\'evy-Leblond method to clarify the naming velocity-space, the velocity-time and space-time contractions. For that we work with the kinematical parameters which are  radius $r$ of the universe, related to the cosmological constant by $r^2=\frac{3}{\Lambda}$, the period $\tau$ of the universe, and the velocity $c$ of light defined by $c=r\tau^{-1}$.\\
In  doing so, the kinematical Lie algebras are found by  the contraction process which consists in keeping one parameter finite and letting the remaining two tend to infinity, their ratio being kept finite.

With section $5$ we introduce the dynamical contractions by first parameterizing the de Sitter Lie algebras by the dynamical parameters mass $m$, compliance $C$ (inverse of stiffness or of Hooke constant or of force constant), and energy, $E_0$. The dynamical parameters and the kinematical parameters are related by $(r^2,\tau^2)=C(E_0,m)$ implying that $E_0=mc^2$.\\
The corresponding contraction consist in letting one of the dynamical parameters go to infinity without constraining the remaining ones, contrary to the kinematical contractions process. The three Bacry-L\'evy-Leblond contractions, i.e. the velocity-space contraction, the velocity-time contraction and the space-time contraction correspond then respectively to an infinite energy $E_0$, an infinite mass $m$ and an infinite compliance $C$. They are the Newtonian limit, the static limit and the flat limit of Dyson \cite{dyson}.

Finally in section $6$, the Kirillov method is used to establish, for each kinematical Lie algebra, a Poisson-Lie algebra and the equations of change that clarify the relationships and differences between the twelve kinematical Lie algebras according the up-down, right-left and frontward-backward contractions (see figure 2).

\section{Inonu-Wigner Contractions of Lie algebras}
We start with a Lie algebra $(\cal{G},\varphi)$ where $\cal{G}$ is a vector space generated by $X_i$ and $\varphi$ is a skew symmetric mapping $\varphi:\cal{G}\times \cal{G}\rightarrow \cal{G}$ defined by $\varphi(X_i,X_j)=X_kC^k_{ij}$ and satisfying the Jacobi identity
\begin{eqnarray}\label{Liealgebra}
\varphi(X_i,\varphi(X_j,X_k))+\varphi(X_j,\varphi(X_k,X_i))+\varphi(X_k,\varphi(X_i,X_j))=0,~\forall X_i,X_j,X_k\in \cal{G}.
\end{eqnarray}
The $C^k_{ij}$ are called the structure constants of the Lie algebra $(\cal{G},\varphi)$. The Jacobi identity shows that a Lie algebra is non associative algebra. \\If the mapping $\psi_{\epsilon}:{\cal{G}}\rightarrow {\cal{G}}$ is singular for a certain value $\epsilon_0$ of $\epsilon$ and if the mapping $\varphi^{\prime}:\cal{G}\times \cal{G}\rightarrow \cal{G}$ is defined by
\begin{eqnarray}
\varphi^{\prime}(X,Y)=\lim_{\epsilon \rightarrow \epsilon_0}\psi^{-1}_{\epsilon}(\psi_{\epsilon}(X),\psi_{\epsilon}(Y))
\end{eqnarray}
then $(\cal{G},\varphi^{\prime})$ is a new Lie algebra called the \textbf{contraction of the Lie algebra} $(\cal{G},\varphi)$ \cite{dooley}.\\
The pioneering contraction method is that of E.Inonu and E.P.Wigner \cite{inonuwigner} which starts with a Lie algebra  $\cal{G}=\cal{H}+\cal{P}$ where $\cal{H}$ is generated by $X_a$, $\cal{P}$ is generated by $X_{\alpha}$; the structure of $\cal{G}$ being a priori given by
\begin{eqnarray*}
\varphi(X_a,X_b)=X_cC^c_{ab}+X_{\gamma}C^{\gamma}_{ab},~\varphi(X_a,X_{\alpha})=X_cC^c_{a\alpha}+X_{\gamma}C^{\gamma}_{a\alpha}~,~\varphi(X_{\alpha},X_{\beta})=X_cC^c_{\alpha\beta}+X_{\gamma}C^{\gamma}_{\alpha\beta}
\end{eqnarray*}
where $a,b,c=1,...,dim(\cal{H})$ and $\alpha,\beta,\gamma =1,...,dim(\cal{P})$.\\
The Inonu-Wigner method uses the parameterized change of basis $\psi_{\epsilon}: (X_a,X_{\alpha})\rightarrow (Y_a,Y_{\alpha})$ defined by  $Y_a=X_a~,~Y_{\alpha}=\epsilon X_{\alpha}$. The structure of the Lie algebra $\cal{G}$ becomes then
\begin{eqnarray*}
\varphi(Y_a,Y_b)=Y_cC^c_{ab}+\epsilon^{-1}Y_{\gamma}C^{\gamma}_{ab},~\varphi(Y_a,Y_{\alpha})=\epsilon Y_cC^c_{a\alpha}+Y_{\gamma}C^{\gamma}_{a\alpha}~,~\varphi(Y_{\alpha},Y_{\beta})=\epsilon^2Y_cC^c_{\alpha\beta}+\epsilon^{-1} Y_{\gamma}C^{\gamma}_{\alpha\beta}
\end{eqnarray*}
In the limit $\epsilon\rightarrow 0$ , the terms $\epsilon^{-1}Y_{\gamma}C^{\gamma}_{ab}$ and $\epsilon^{-1} Y_{\gamma}C^{\gamma}_{\alpha\beta}$ diverge. A limit will exist if only if the structure constants $C^{\gamma}_{ab}$ and $C^{\gamma}_{\alpha\beta}$ vanish. Hence to get a Inonu-Wigner contraction, $\cal{H}$ must be a subalgebra of $\cal{G}$.  The structure of the contracted Lie algebra is then
\begin{eqnarray}\label{contracted}
\varphi^{\prime}(Y_a,Y_b)=Y_cC^c_{ab}~,~\varphi^{\prime}(Y_a,Y_{\alpha})=Y_{\beta}C^{\beta}_{a\alpha}~,~\varphi^{\prime}(Y_{\alpha},Y_{\beta})=0
\end{eqnarray}
The Lie algebra $(\cal{G},\varphi^{\prime})$ defined by (\ref{contracted}) is a Inonu-Wigner contraction of the mother Lie algebra $(\cal{G},\varphi)$ with respect to the Lie subalgebra $\cal{H}$. It is a semi-direct sum of $(\cal{H},\varphi^{\prime})$ and the abelian Lie algebra $(\cal{P},\varphi^{\prime})$.
\section{Possible kinematical Lie algebras \`a la L\'evy-Leblond}
According to H.Bacry and J.M.L\'evy-Leblond \cite{jmll}, a kinematical group is a space-time transformation group which keeps laws of physics invariant. Due to the assumptions of space isotropy, space-time homogeneity and existence of inertial transformations, a kinematical group is a ten dimensional Lie group whose Lie algebra is generated by three rotation generators $J_i$ (isotropy of space), three space translation generators $P_i$ (homogeneity of space), a time translation generator $H$ (homogeneity of time) and three inertial transformation generators $K_i$. Following Bacry and L\'evy-Leblond \cite{jmll}, A.Ngendakumana and coauthors \cite{ancilla1}  have shown that under some mathematical physics assumptions  only twelve kinematical Lie algebras exist. Their Lie algebraic structures have in common the Lie brackets defining the adjoint representation of the rotation generators
\begin{eqnarray*}
[J_i,J_j]=J_k\epsilon^k_{ij},~[J_i,K_j]=K_k\epsilon^k_{ij},~[J_i,P_j]=P_k\epsilon^k_{ij},~[J_i,H]=0
\end{eqnarray*}
The remaining Lie brackets are given by the Table I \cite{ancilla1}. The ParaPoincar\'e Lie algebra ${\cal{P}},_+$ which is isomorphic to the Euclidean Lie algebra ${\cal{E}}(4)$ where the "translations" generated by $K_i$ and $H$ form an abelian Lie subalgebra does not appear in the list of kinematical ones by Bacry and L\'evy-Leblond \cite{jmll}. The argument is that the inertial transformations are compact. However they are noncompact and only space translations are compact.\\Using the Inonu-Wigner contraction method \cite{inonuwigner}, Bacry and L\'evy-Leblond \cite{jmll} have established that these Lie algebras are approximations of the de Sitter Lie algebras. Their links are summarized by the contractions scheme (see Figure $1$ on page $1610$ of \cite{jmll}). We will refer to these nomenclature in the next two sections.

\section{Kinematical Lie algebras \`a la L\'evy-Leblond revisited}
We propose to recover the Bacry-L\'evy-Leblond contractions scheme by using the kinematical parameters $r$, $\tau$ and $c$ which are respectively the radius of universe, the period of the universe and speed of light. We first introduce the de Sitter Lie algebras $dS_{\pm}$ as isomorphic to the pseudo-orthogonal Lie algebras $O_{\pm}(5)$, i.e. that $dS_+(3)$ [$dS_-(3)$] is isomorhic to $O(1,4)$ [$O(2,3)$] Lie algebra. The aim of this section is to better clarify  velocity-space contractions, velocity-time contractions and space-time contractions of Bacry and L\'evy-Leblond \cite{jmll}.
\subsection{The Lie algebras ${\cal{O}}_{\pm}(5)$}
Let $V$ be a five dimensional manifold equipped with the metric
\begin{eqnarray}\label{metric}
ds^2=\delta_{ij}dx^idx^j-(dx^4)^2\pm(dx^5)^2\equiv \eta_{ab}dx^adx^b,~~i,j=1,2,3
\end{eqnarray}
where the dimension of the $x^a$ is that of length. The matrix elements $\eta_{ab}$ form the diagonal matrix $\eta_{\pm}=diag(I_{3\times 3},-1,\pm 1)$.
The isometry group $G_{\pm}$ of the metric (\ref{metric}) is the group of real square matrices $g$ with order five satisfying $g^t\eta_{\pm}g=\eta_{\pm}$.
The Lie group $SO_0(4,1)$ is the connected component of $G_{+}$ while $SO_0(3,2)$ is the connected component of $G_-$. The Lie algebra ${\cal{O}}_{\pm}(5)$ is the set of the real square matrices $X$ of order $5$ satisfying $^tX\eta_{\pm}+\eta_{\pm}X=0$. We easily verify that $X=J_k\theta^k+A_k\alpha^k+B_k\beta^k+\gamma \Gamma,~~~k=1,2,3,~$ is the dimensionless matrix
\begin{eqnarray}\label{ofivematrix}
X=\left(
\begin{array}{ccc}
\epsilon_{kj}^i\theta^k&\alpha^i&\beta^i\\\alpha_j&0&\gamma\\\mp\beta_j&\pm \gamma&0
\end{array}
\right)
\end{eqnarray}
and that $(J_k,~A_k~B_k,~\Gamma)$ is a basis of ${\cal{O}}_{\pm}(5)$. The Lie algebra ${\cal{O}}_{\pm}(5)$ structure is defined by the Lie brackets
\begin{eqnarray}\label{rotations}
[J_i,J_j]=J_k\epsilon^k_{ij},~[J_i,A_j]=A_k\epsilon^k_{ij},~[J_i,B_j]=B_k\epsilon^k_{ij},~[J_i,\Gamma]=0
\end{eqnarray}
\begin{eqnarray}\label{boosts}
[A_i,A_j]=- J_k\epsilon^k_{ij},~[A_i,B_j]=\Gamma \delta_{ij},~[A_i,\Gamma]=B_i
\end{eqnarray}
\begin{eqnarray}\label{translations}
[B_i,B_j]=\pm J_k\epsilon^k_{ij},~[B_i,\Gamma]=\pm A_i
\end{eqnarray}
\subsection{The de Sitter Lie algebras}
Let $K_i=\frac{1}{c}A_i$, $P_i=\frac{1}{r}B_i$ and $H=\frac{1}{\tau}\Gamma$ where $\sigma=\frac{1}{c}$ is a slowness, $\kappa=\frac{1}{r}$ is a curvature while $\omega=\frac{1}{\tau}$ is a frequency. In the new basis $(J_k,~K_k~P_k,~H)$ of  ${\cal{O}}_{\pm}(5)$ the matrix $X$ above becomes
\begin{eqnarray}\label{desittermatrix}
X=\left(
\begin{array}{ccc}
\epsilon_{kj}^i\theta^k&\frac{v^i}{c}&\frac{x^i}{r}\\\frac{v_j}{c}&0&\frac{t}{\tau}\\\mp\frac{x_j}{r}&\pm \frac{t}{\tau}&0
\end{array}
\right)
\end{eqnarray}
where $\alpha^i=\frac{v^i}{c}~,\beta^i=\frac{x^i}{r}~,\gamma=\frac{t}{\tau}$. Hence the parameters associated with $K_i$ ,$P_i$ and $H$ have velocity, length and time as respective physical dimension. The Lie brackets (\ref{rotations}), (\ref{boosts}) and (\ref{translations}) become then
\begin{eqnarray}\label{dsrotations1}
[J_i,J_j]=J_k\epsilon^k_{ij},~[J_i,K_j]=K_k\epsilon^k_{ij},~[J_i,P_j]=P_k\epsilon^k_{ij},~[J_i,H]=0
\end{eqnarray}
\begin{eqnarray}\label{dsboosts1}
[K_i,K_j]=-\frac{1}{c^2} J_k\epsilon^k_{ij},~[K_i,P_j]=\frac{\tau}{c r}H\delta_{ij},~[K_i,H]=\frac{r}{c \tau}P_i
\end{eqnarray}
\begin{eqnarray}\label{dstranslations1}
[P_i,P_j]=\pm\frac{1}{r^2} J_k\epsilon^k_{ij},~[P_i,H]=\pm \frac{c}{r \tau}K_i
\end{eqnarray}
Let us now study the limits of the de Sitter Lie algebras as the constants tend to infinity. Normally the three constants are constrained by $c=r\tau^{-1}$. However, we ignore for a moment. We use it at the end of the section to show that our way of doing has recovered the results of table $1$. It is first of all evident that (\ref{dsrotations1}) does not change. We are then only interested in the behavior of (\ref{dsboosts1}) and (\ref{dstranslations1}).
\subsection{The Newton-Hooke, Poincar\'e and Para-Poincar\'e Lie algebras}
In this section we look for the limits of the de Sitter Lie algebras as two of the constants tend to infinity while their ratio is kept finite.
\subsubsection{Newton-Hooke Lie algebras}
 We verify that the limits of (\ref{dsboosts1}) and (\ref{dstranslations1}),  as the speed $c$ and the radius $r$ tend to infinity while their ratio $\frac{r}{c}$ and $\tau$ are kept finite,  are
\begin{eqnarray}\label{dsboosts2}
[K_i,K_j]=0,~[K_i,P_j]=0,~[K_i,H]=\frac{r}{c\tau}P_i
\end{eqnarray}
\begin{eqnarray}\label{dstranslations2}
[P_i,P_j]=0,~[P_i,H]=\pm \frac{c}{r\tau}K_i
\end{eqnarray}
The Lie brackets (\ref{dsrotations1}), (\ref{dsboosts2}) and (\ref{dstranslations2}) define the Newton-Hooke Lie algebra ${\cal{NH}}_{\pm}$.
\subsubsection{Poincar\'e Lie algebra}
 If the period $\tau$ and the radius $r$ tend to infinity while their ratio $\frac{r}{\tau}$ and $c$ are kept finite, the brackets (\ref{dsboosts1}) and (\ref{dstranslations1}) become
\begin{eqnarray}\label{dsboosts3}
[K_i,K_j]=-\frac{1}{c^2}J_k\epsilon^k_{ij},~[K_i,P_j]=\frac{\tau}{rc}H\delta_{ij},~[K_i,H]=\frac{r}{c\tau}P_i
\end{eqnarray}
\begin{eqnarray}\label{dstranslations3}
[P_i,P_j]=0,~[P_i,H]=0
\end{eqnarray}
The Lie brackets (\ref{dsrotations1}), (\ref{dsboosts3}) and (\ref{dstranslations3}) define the Poincar\'e Lie algebra ${\cal{P}}$.
\subsubsection{Para-Poincar\'e Lie algebras}
 Similarly if the speed $c$ and the time $\tau$ tend to infinity while their ratio $\frac{c}{\tau}$ and $r$ are kept finite then the Lie brackets (\ref{dsboosts1}) and (\ref{dstranslations1}) become
\begin{eqnarray}\label{dsboosts4}
[K_i,K_j]=0,~[K_i,P_j]=\frac{\tau}{rc}H\delta_{ij},~[K_i,H]=0
\end{eqnarray}
\begin{eqnarray}\label{dstranslations4}
[P_i,P_j]=\pm\frac{1}{r^2}J_k\epsilon^k_{ij},~[P_i,H]=\pm\frac{c}{\tau r}K_i
\end{eqnarray}
The Lie brackets (\ref{dsrotations1}), (\ref{dsboosts4}) and (\ref{dstranslations4}) define  the Para-Poincar\'e Lie algebra ${\cal{P}}_{\pm}$.\\
We then notice that the Newton-Hooke Lie algebras, the Poincar\'e Lie algebra and the Para-Poincar\'e Lie algebras are respectively the velocity-space, space-time and velocity-time contractions of the de Sitter Lie algebras as in \cite{jmll}.
\subsection{The Galilei, Para-Galilei and Carroll Lie algebras}
\subsubsection{Galilei Lie algebra}
The limit of the Lie brackets (\ref{dsboosts2}) and (\ref{dstranslations2}) as the radius $r$ and the period $\tau$ tend to infinity while $\frac{r}{\tau}$ and $c$ are kept finite and the limit (\ref{dsboosts3}) and (\ref{dstranslations3}) as the radius $r$ and the speed $c$ tend to infinity while $\frac{c}{r}$ and $\tau$ are kept finite are the same, i.e.
\begin{eqnarray}\label{dsboosts5}
[K_i,K_j]=0,~[K_i,P_j]=0,~[K_i,H]=\frac{r}{c\tau} P_i
\end{eqnarray}
\begin{eqnarray}\label{dstranslations5}
[P_i,P_j]=0,~[P_i,H]=0
\end{eqnarray}
The Lie brackets (\ref{dsrotations1}), (\ref{dsboosts5}) and (\ref{dstranslations5}) define the Galilei Lie algebra $\cal{G}$.
\subsubsection{Para-Galilei Lie algebras}
The limit of the Lie brackets (\ref{dsboosts2}) and (\ref{dstranslations2}) as the speed $c$ and the period $\tau$ tend to infinity while $\frac{c}{\tau}$ and $r$ are kept finite and the limit (\ref{dsboosts4}) and (\ref{dstranslations4}) as the radius $r$ and the speed $c$ tend to infinity while $\frac{c}{r}$ and $\tau$ are kept finite are the same, i.e.
\begin{eqnarray}\label{dsboosts6}
[K_i,K_j]=0,~[K_i,P_j]=0,~[K_i,H]=0
\end{eqnarray}
\begin{eqnarray}\label{dstranslations6}
[P_i,P_j]=0,~[P_i,H]=\pm\frac{c}{r\tau} K_i
\end{eqnarray}
The Lie brackets (\ref{dsrotations1}), (\ref{dsboosts6}) and (\ref{dstranslations6}) define the Para-Galilei Lie algebras $\cal{G}_{\pm}$.
\subsubsection{Carroll Lie algebra}
The limit of the Lie brackets (\ref{dsboosts3}) and (\ref{dstranslations3}) as the speed $c$ and the period $\tau$ tend to infinity while $\frac{c}{\tau}$ is kept finite and the limit (\ref{dsboosts4}) and (\ref{dstranslations4}) as the radius $r$ and the period $\tau$ tend to infinity while $\frac{r}{\tau}$ is kept finite are the same, i.e.
\begin{eqnarray}\label{dsboosts7}
[K_i,K_j]=0,~[K_i,P_j]=\frac{\tau}{rc} H\delta_{ij},~[K_i,H]=0
\end{eqnarray}
\begin{eqnarray}\label{dstranslations7}
[P_i,P_j]=0,~[P_i,H]=0
\end{eqnarray}
The Lie brackets (\ref{dsrotations1}), (\ref{dsboosts7}) and (\ref{dstranslations7}) define the Carroll Lie algebra ${\cal{C}}$.\\
Hence the Galilei, the Para-Galilei, the Carroll Lie algebras are respective contractions of the Newton-Hooke or Poincar\'e Lie algebras, the Newton-Hooke or the Para-Poincar\'e Lie algebras, the Poincar\'e or the Para-Poincar\'e Lie algebras respectively.
\subsection{The Static Lie algebra}
The limit of the Lie brackets (\ref{dsboosts5}) and (\ref{dstranslations5}) as the speed $c$ and the period $\tau$ tend to $\infty$ while $\frac{c}{\tau}$ and $r$ are kept finite ,the limit (\ref{dsboosts6}) and (\ref{dstranslations6}) as the radius $r$ and the period $\tau$ tend to $\infty$ while $\frac{r}{\tau}$ and $c$ are kept finite and the limit of the Lie brackets (\ref{dsboosts7}) and (\ref{dstranslations7}) as the speed $c$ and the radius $r$ tend infinity while $\frac{c}{r}$  and $\tau$ are kept finite are the same; i.e
\begin{eqnarray}\label{dsboosts8}
[K_i,K_j]=0,~[K_i,P_j]=0,~[K_i,H]=0
\end{eqnarray}
\begin{eqnarray}\label{dstranslations8}
[P_i,P_j]=0,~[P_i,H]=0
\end{eqnarray}
The Lie brackets (\ref{dsrotations1}), (\ref{dsboosts8}) and (\ref{dstranslations8}) define the Static Lie algebra $\cal{S}$.\\When the constraint $c=r\tau^{-1}$ is taken in account, the Lie brackets in the table I are recovered.\\These approximations through kinematical parameters are summarized in the following cube (see figure $1$).  On the cube, the {\color{green} horizontal arrows} represent the contractions as {$c,\tau \rightarrow \infty$, $\frac{c}{\tau}$ and $r$ finite} ( velocity-time contractions ), the {\color{red} vertical arrows} represent the contractions as $c,r \rightarrow \infty$, $\frac{r}{c}$ and $\tau$ finite (velocity-space contractions) and the {\color{blue} oblique arrows} represent the contractions as $r,\tau \rightarrow \infty$, $\frac{r}{\tau}$ and $c$ finite (space-time contractions).

\section{Dynamical contractions of the de Sitter Lie algebras}
The main contribution of this paper is to show that there are three parameters characterizing the de Sitter Lie algebras such that the three kind of Bacry- L\'evy-Leblond contractions be obtained by letting one parameter tend to infinity without constraining the remaining ones. Instead of working with the kinematical parameters the radius of universe $r$, the speed of light $c$ and the period $\tau$ related by (\ref{ligthspeed}), we introduce the dynamical parameters compliance $C$, mass $m$ and energy $E_0$. These dynamical parameters enter the de Sitter Lie algebras structure by replacing the boost generators $K_i$ by the momentum generators $Q_i=\frac{1}{m}K_i$, $m$ being a mass and by defining the compliance $C$ and the energy $E_0$ respectively by $C=\frac{\tau^2}{m}$ and $E_0=mc^2$.
 \subsection{Three parameters Lie algebras} The de Sitter Lie algebras $dS_{\pm}$ are then defined in the basis $(J_i,Q_i,P_i,H)$, by the Lie brackets
\begin{eqnarray}\label{dsrotations9}
[J_i,J_j]=J_k\epsilon^k_{ij},~[J_i,Q_j]=Q_k\epsilon^k_{ij},~[J_i,P_j]=P_k\epsilon^k_{ij},~[J_i,H]=0
\end{eqnarray}
\begin{eqnarray}\label{dsboosts9}
[Q_i,Q_j]=-\frac{1}{mE_0} J_k\epsilon^k_{ij},~[Q_i,P_j]=\frac{1}{E_0}H\delta_{ij},~[Q_i,H]=\frac{1}{m}P_i
\end{eqnarray}
\begin{eqnarray}\label{dstranslations9}
[P_i,P_j]=\pm\frac{1}{CE_0} J_k\epsilon^k_{ij},~[P_i,H]=\pm \frac{1}{C}Q_i
\end{eqnarray}
The de Sitter Lie algebras $dS_{\pm}$ are then characterized by the three dynamical parameters $m$, $C$ and $E_0$.
\subsection{Two parameters Lie algebras}
When the energy $E_0$ tends to infinity, the compliance $C$ tends to infinity and the mass tends to infinity, the Lie brackets (\ref{dsrotations9}), (\ref{dsboosts9}) and (\ref{dstranslations9}) tend respectively to the Lie brackets defining the Newton-Hooke Lie algebras ($NH_{\pm}$), the Poincar\'e Lie algebra $P$ and the Para-Poincar\'e Lie algebras $P_{\pm}$. The Newton-Hooke Lie algebras are then characterized by a mass $m$ and a compliance $C$ related by the frequency $\omega=\frac{1}{\sqrt{mC}}$ (time $\tau=\sqrt{mC}$), the Poincar\'e Lie algebra is characterized by a mass and an energy related by the speed $c=\sqrt{\frac{E_0}{m}}$ (slowness $s=\sqrt{\frac{m}{E_0}}$) and the Para-Poincar\'e Lie algebras are characterized by an energy and a compliance related by the curvature $\kappa=\frac{1}{\sqrt{CE_0}}$ (the radius $r=\sqrt{CE_0}$).
\subsection{One parameter Lie algebras}
The Lie brackets which define the Galilei Lie algebra $G$ are obtained from those defining the Poincar\'e algebra as the energy tends to infinity or from those defining the Newton-Hooke Lie algebras as the compliance tends to infinity; the Lie brackets defining the Carroll Lie algebra $C$ are obtained from those defining the Poincar\'e Lie algebra as the mass tends to infinity or from those defining the Para-Poincar\'e Lie algebras as the compliance tends to infinity; the Lie brackets defining the Para-Galilei Lie algebras $G_{\pm}$ are obtained from those defining the Para-Poincar\'e Lie algebras as the energy tends to infinity or from those defining the Newton-Hooke Lie algebras as the mass tends to infinity. The Galilei Lie algebra is then characterized by a mass, the Carroll Lie algebra is characterized by an energy while the Para-Galilei Lie algebras are characterized by a compliance. We can say that the mass $m$ is galilean, the compliance $C$ is para-galilean and the energy $E_0$ is carrollian.
\subsection{Zero parameters Lie algebra}
The zero parameters Lie algebra is the static Lie algebra which is obtained from the Galilei Lie algebra as the mass tends to infinity, from the Carroll Lie algebra as the energy tends to infinity or from the Para-Galilei Lie algebra as the compliance tends to infinity.\\All the Lie brackets defining these Lie algebras have in common the Lie brackets
\begin{eqnarray*}
[J_i,J_j]=J_k\epsilon^k_{ij},~[J_i,Q_j]=Q_k\epsilon^k_{ij},~[J_i,P_j]=P_k\epsilon^k_{ij},~[J_i,H]=0
\end{eqnarray*}
the others are summarized in the table II while the limiting process is given by the figure $2$ where the horizontal arrows represent the contractions as {\color{green}$m \rightarrow \infty$} (static limit), the vertical arrows represent the contractions as {\color{red}$E_0 \rightarrow \infty$} (Newtonian limit) and the oblique arrows represent the contractions as {\color{blue}$C \rightarrow \infty$} (flat limit). If we use coordinates $(\frac{1}{m},\frac{1}{C},\frac{1}{E_0})$, the kinematical Lie algebras costitute then the cube below (figure 2). The table III give a comparison of kinematical Lie algebras distribution obtained through the dynamical contraction process above with that obtained by the kinematical contraction process as given by MacRae \cite{mcrae}.
  \section{A glance at the physics associated to the kinematical Lie algebras}
 Let us have a look at the physics associated to the kinematical Lie algebras in function of the three dynamical parameters.
 \subsection{Poisson brackets}
 We know that the Poisson bracket of two functions defined on the dual ${\cal{G}}^*$ of any Lie algebra $\cal{G}$ is defined by
 \begin{eqnarray}\label{Kirillovpoissonbrackets}
 \{f,g\}=K_{ij}(a)\frac{\partial f}{\partial a_i}\frac{\partial g}{\partial a_j}
 \end{eqnarray}
where $a_i$ are the coordinates on ${\cal{G}}^*$ and $K_{ij}(a)=-a_kC^k_{ij}$ are the matrix elements of the Kirillov form.

 Let the general element of the dual of a kinematical Lie algebra be $j_kJ^{k*}+q_kQ^{k*}+\pi_kP^{k*}+EH^*$ where $j_k$ are the components of the angular momentum, $\pi_k$ are the components of the linear momentum, $q_k$ are the components of the position while $E$ is an energy. It follows from the Lie brackets (see previous section) defining the kinematical Lie algebras in function of $m$,$E_0$ and $C$ that the kinematical Poisson-Lie algebras are defined by
  \begin{eqnarray}\label{poissonrotations}
\{j_i,j_j\}=-j_k\epsilon^k_{ij},~\{j_i,q_j\}=-q_k\epsilon^k_{ij},~\{j_i,\pi_j\}=-\pi_k\epsilon^k_{ij},~\{j_i,E\}=0
\end{eqnarray}
 and the other Poisson brackets given by the table $V$.
\subsection{Equations of Change}
We rewrite (\ref{Kirillovpoissonbrackets}) as $\{f,g\}=X_f(g)$ where the vector field $X_f$ is defined by
\begin{eqnarray}
X_f=K_{ij}(a)\frac{\partial f}{\partial a_i}\frac{\partial }{\partial a_j}
\end{eqnarray}
and verifies $[X_f,X_g]=X_{\{f,g\}}$. It is known that the mapping $\rho$ defined by $\rho(f)=X_f$ is a realization of the Lie algebra ${\cal{G}}$.\\ The de Sitter Lie algebras $d{\cal{S}}_{\pm}$ is the realized by the vector fields
\begin{eqnarray}\label{desitterhamfieldrot}
X_{j_i}=-j_k\epsilon^k_{ij}\frac{\partial}{\partial j_j}-q_k\epsilon^k_{ij}\frac{\partial}{\partial q_j}-\pi_k\epsilon^k_{ij}\frac{\partial}{\partial \pi_j}
\end{eqnarray}
\begin{eqnarray}\label{desitterhamfieldmomentum}
X_{q_i}=q_k\epsilon^k_{ij}\frac{\partial}{\partial j_j}+\frac{j_k}{mE_0}\epsilon^k_{ij}\frac{\partial}{\partial q_j}-\frac{E}{E_0}\frac{\partial}{\partial \pi_i}-\frac{\pi_i}{m}\frac{\partial}{\partial E}
\end{eqnarray}
\begin{eqnarray}\label{desitterhamfieldspace}
X_{\pi_i}=\pi_k\epsilon^k_{ij}\frac{\partial}{\partial j_j}+\frac{E}{E_0}\frac{\partial}{\partial q_i}\mp\frac{j_k}{CE_0}\epsilon^k_{ij}\frac{\partial}{\partial \pi_j}\mp\frac{q_i}{C}\frac{\partial}{\partial E}
\end{eqnarray}
\begin{eqnarray}\label{desitterhamfieldtime}
X_E=\frac{\pi_i}{m}\frac{\partial}{\partial q_i}\pm\frac{q_i}{C}\frac{\partial}{\partial \pi_i}~~~~~~~~~~~~~~~
\end{eqnarray}
The realizations of the other kinematical Lie algebras are obtained from (\ref{desitterhamfieldrot}) to (\ref{desitterhamfieldtime}) through the dynamical contraction process defined in the previous section. No need to make them explicit  here.\\If $X_f$ is the generating function of the one parameter diffeomorphism $\Phi_s:{\cal{G}}^*\rightarrow {\cal{G}}^*$ and if $X_g$ is the generating function of the one parameter diffeomorphism $\Phi_{\lambda}:{\cal{G}}^*\rightarrow {\cal{G}}^*$, then the equation of change of the function $g$ with respect to $s$ is $\frac{dg}{ds}=X_f(g)$ while the equation of change of $f$ with respect $\lambda$ is $\frac{df}{d\lambda}=X_g(f)$, $\frac{dg}{ds}=-\frac{df}{d\lambda}$. We illustrate the change equations by using the time $t$, the longitude $\varphi=\theta^3$, the altitude $z=x^3$ and the corresponding momentum up $p_z$. Note that equations of change with respect the time $t$ are called motion (evolution) equations.\\ For each parameter $s$, let define $F_s$ as $F_s=\{f\in {\cal{G}}^*:\frac{df}{ds}=0\}$ and let $V_s={\cal{G}}^*/F_s$ be the variables submanifold of ${\cal{G}}^*$. The equations of change describe how the coordinates on $V_s$ change with respect the ad hoc parameter. Note that as $\Phi_s \circ \Phi_t =\Phi_{s+t}$, the change parameters must be additive. It is true for the longitudinal angle. We show in the appendix that it is also true for the momentum parameter $p^i$, the space translation $x^i$ and the time translation parameter $t$ appearing in the $dS_{\pm}$ Lie algebra element $X=J_k\theta^k+Q_kp^k+P_kx^k+Ht$. We use the one spatial Poincar\'e, Para-Poincar\'e and Newton-Hooke, Lie algebras to respectively associate a non additive boost to momentum, a non additive force to space translation and to time translation a non additive...\fbox{\large WHAT ???} The equations of change are given in the table $V$.\\
In this table the greec indices take the value $1$ and $2$ while the latin indices take the values $1,2$ and $3$. The equations of the form $\frac{df}{ds}=0$ do not appear in the table. Moreover the first column of the table contains the dimensions of $V_s$, the second one indicates the first order differential equations, the third one the second order differential equation when available, finally the last one shows the corresponding kinematical Lie algebras paired as mother-daughter in the parental relations (see figure $2$) up-down for the variations with respect time (energy $E_0$ is absent in the equations), front-backward for the variations with respect the altitude $z$ (mass $m$ is absent in the equations) and right-left for the variations with respect the momentum up $p$ ( compliance $C$ is absent in the equations). We also notice that $V_{\varphi}=V_{\varphi}(j_{\mu})\oplus V_{\varphi}(q_{\mu})\oplus V_{\varphi}(\pi_{\mu})$  (i.e. $6=2+2+2$ as sum of dimensions) under the differential rotation operator $\frac{d}{d\varphi}$. Note that $V_{\varphi}(j_{\mu})$ means that the components of $j$ form an irreducible entity under the differential operator $\frac{d}{d\varphi}$. Also $V_t$ is irreducible under the differential time operator $\frac{d}{dt}$ in the de Sitter and Newton-Hooke case, that $V_z=V_z(j_{\mu},\pi_{\mu})\oplus V_z(q,E)$ (i.e. $6=4+2$) is irreducible under the differential altitude operator $\frac{d}{dz}$ in the de Sitter and Para-Poincar\'e cases, and finally that $V_p=V_p(j_{\mu},q_{\mu})\oplus V_p(\pi,E)$ (i.e.$6=4+2$) is irreducible under the differential momentum up operator $\frac{d}{dp}$ in the de Sitter and Poincar\'e cases. The reader can also verify that the three dimensional manifolds are irreducible under the operator $\frac{d}{dt}$. They are direct sums (i.e. 3=2+1) under the operators $\frac{d}{dz}$ and $\frac{d}{dp}$. Finally the two dimensional ones are irreducible and correspond to the pairs containing the static Lie algebra as a daughter. Note also that all the ten coordinates on ${\cal{G}}^*$ are constant in the Carroll and static Lie algebras cases.
\section{Conclusion}
In this paper we have shown how to obtain  straightfully all  kinematical Lie algebras from the de Sitter Lie algebras through  contraction  using dynamical parameters mass $m$, compliance $C$ and energy $E_0$. We had a little glance at the physics associated to each kinematical Lie algebra. We noticed that $V_{\varphi}=V_{\varphi}(j_{\mu})\oplus V_{\varphi}(q_{\mu})\oplus V_{\varphi}(\pi_{\mu})$  (i.e. $6=2+2+2$) under the differential rotation operator $\frac{d}{d\varphi}$, where $V_s(j_{\mu})$ means the components of $j$ form an irreducible entity under the differential operator $\frac{d}{ds}$. Also $V_t$ is irreducible under the differential time operator $\frac{d}{dt}$ in the de Sitter and Newton-Hooke case, that $V_z=V_z(j_{\mu},\pi_{\mu})\oplus V_z(q,E)$ (i.e. $6=4+2$) under the differential altitude operator $\frac{d}{dz}$ in the de Sitter and Para-Poincar\'e cases and finally that $V_p=V_p(j_{\mu},q_{\mu})\oplus V_p(\pi,E)$ (i.e.$6=4+2$) under the differential momentum operator $\frac{d}{dp}$ in the de Sitter and Poincar\'e cases.\\
We notice from table $IV$ that positions do not commute in the de Sitter and Poincar\'e cases, that linear momenta do not commute in de Sitter and Para-Poicare cases and that the uncertainty - like relation $\{\pi_i,q_j\}=\frac{E}{E_0}\delta_{ij}$ occurs in  finite energy (relative time) groups.
\begin{table}[htbp]
\center{}
\begin{tabular}{|c|c|c|c|c|c|c|c|}
\hline
Lie symbol&Lie algebra Name& $[K_i,H]$&$[K_i,K_j]$&$[K_i,P_j]$&$[P_i,P_j]$& $[P_i,H]$\\
\hline
 $dS_{\pm}$& $de~Sitter$ &$P_i$ & $-\frac{1}{c^2}J_k\epsilon^k_{ij}$ & $\frac{1}{c^2}H\delta_{ij}$ & $\pm \frac{1}{r^2}J_k\epsilon^k_{ij}$& $\pm \frac{1}{\tau^2} K_i$ \\
 \hline
  $P$& $Poincare$ & $P_i$ & $-\frac{1}{c^2}J_k\epsilon^k_{ij}$ & $\frac{1}{c^2}H\delta_{ij}$ & $0$& $0$ \\
  $NH_{\pm}$& $Newton-Hooke$ & $P_i$ & $0$ & $0$ & $0$& $\pm \frac{1}{\tau^2} K_i$ \\
  $P_{\pm}$& $Para~Poincare$ & $0$ & $0$ & $\frac{1}{c^2}H\delta_{ij}$ & $\pm \frac{1}{r^2}J_k\epsilon^k_{ij}$& $\pm \frac{1}{\tau^2} K_i$ \\
  \hline
$G$& $Galilei$ & $P_i$ & $0$ & $0$ & $0$& $0$ \\
$G_{\pm}$& $Para-Galilei$ & $0$ & $0$ & $0$ & $0$& $\pm \frac{1}{\tau^2} K_i$ \\
$C$& $Carroll$ & $0$ & $0$ & $\frac{1}{c^2}H\delta_{ij}$ & $0$& $0$ \\
\hline
 $S$& $Static$ & $0$ & $0$ & $0$ & $0$& $0$ \\
 \hline
\end{tabular}
\caption{The kinematical Lie algebras in term of $c$,$r$ and $\tau$}
\end{table}
\pagebreak
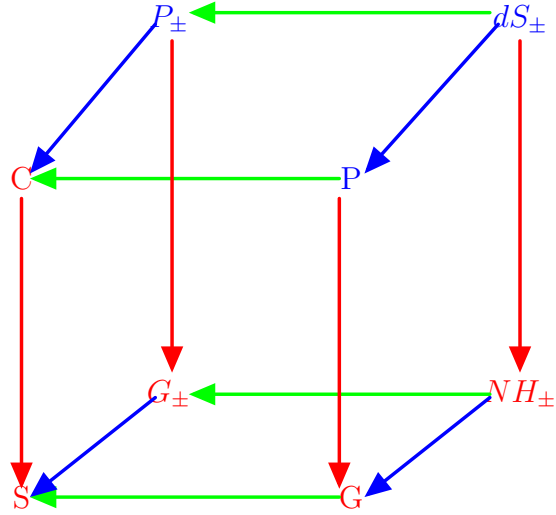
\begin{figure}[h!]
\begin{center}
\definecolor{qqqqff}{rgb}{0,0,1}
\begin{tikzpicture}[line cap=round,line join=round,>=triangle 45,x=1.0cm,y=1.0cm,scale=1.1]
\draw[color=red] (1.5,1) node {S};
\draw[color=red] (1.5,4.84) node {C};
\draw[color=red] (5.44,1) node {G};
\draw[color=qqqqff] (5.44,4.84) node {P};
\draw[color=red] (3.26,2.24) node {$G_{\pm}$};
\draw[color=qqqqff] (3.26,6.76) node {$P_{\pm}$};
\draw[color=red] (7.46,2.24) node {$NH_{\pm}$};
\draw[color=qqqqff] (7.46,6.76) node {$dS_{\pm}$};
{\color{green}\draw[ ->,line width=1.3pt] (5.3,1) -- (1.6,1);}
{\color{green}\draw[ ->,line width=1.3pt] (7.1,2.24) -- (3.5,2.24);}
{\color{green}\draw[ ->,line width=1.3pt] (5.3,4.84) -- (1.6,4.84);}
{\color{green}\draw[ ->,line width=1.3pt] (7.1,6.84) -- (3.5,6.84);}
{\color{red}\draw[ ->,line width=1.3pt] (1.5,4.6) -- (1.5,1.1);}
{\color{red}\draw[ ->,line width=1.3pt] (5.3,4.6) -- (5.3,1.1);}
{\color{red}\draw[ ->,line width=1.3pt] (3.3,6.5) -- (3.3,2.5);}
{\color{red}\draw[ ->,line width=1.3pt] (7.46,6.5) -- (7.46,2.5);}
{\color{blue}\draw[ ->,line width=1.3pt] (3.1,2.2) -- (1.6,1);}
{\color{blue}\draw[ ->,line width=1.3pt] (7.1,2.2) -- (5.6,1);}
{\color{blue}\draw[ ->,line width=1.3pt] (3.1,6.7) -- (1.6,4.9);}
{\color{blue}\draw[ ->,line width=1.3pt] (7.2,6.7) -- (5.6,4.9);}
\end{tikzpicture}
\end{center}
\caption{Contractions in terms of $c$, $r$ and $\tau$}
\end{figure}
\pagebreak
\begin{table}[htbp]
\center{}
\begin{tabular}{|c|c|c|c|c|c|c|c|}
\hline
Lie symbol&Lie algebra Name& $[Q_i,H]$&$[Q_i,Q_j]$&$[Q_i,P_j]$&$[P_i,P_j]$& $[P_i,H]$\\
\hline
 $dS_{\pm}$& $de~Sitter$ &$\frac{1}{m}P_i$ & $-\frac{1}{mE_0}J_k\epsilon^k_{ij}$ & $\frac{1}{E_0}H\delta_{ij}$ & $\pm \frac{1}{CE_0}J_k\epsilon^k_{ij}$& $\pm \frac{1}{C} Q_i$ \\
 \hline
  $P$& $Poincare$ & $\frac{1}{m}P_i$ & $-\frac{1}{mE_0}J_k\epsilon^k_{ij}$ & $\frac{1}{E_0}H\delta_{ij}$ & $0$& $0$ \\
  $NH_{\pm}$& $Newton-Hooke$ & $\frac{1}{m}P_i$ & $0$ & $0$ & $0$& $\pm \frac{1}{C} Q_i$ \\
  $P_{\pm}$& $Para~Poincare$ & $0$ & $0$ & $\frac{1}{E_0}H\delta_{ij}$ & $\pm \frac{1}{CE_0}J_k\epsilon^k_{ij}$& $\pm \frac{1}{C} Q_i$ \\
  \hline
$G$& $Galilei$ & $\frac{1}{m}P_i$ & $0$ & $0$ & $0$& $0$ \\
$G_{\pm}$& $Para-Galilei$ & $0$ & $0$ & $0$ & $0$& $\pm \frac{1}{C} Q_i$ \\
$C$& $Carroll$ & $0$ & $$ & $\frac{1}{E_0}H\delta_{ij}$ & $0$& $0$ \\
\hline
 $S$& $Static$ & $0$ & $0$ & $0$ & $0$& $0$ \\
 \hline
\end{tabular}
\caption{The kinematical Lie algebras in term of the mass, the compliance and the energy}
\end{table}
\pagebreak
\begin{figure}[h!]
\begin{center}
\definecolor{qqqqff}{rgb}{0,0,1}
\begin{tikzpicture}[line cap=round,line join=round,>=triangle 45,x=1.0cm,y=1.0cm,scale=1.1]
\draw[color=red] (1.5,1) node {G};
\draw[color=red] (1.5,4.84) node {P};
\draw[color=red] (5.44,1) node {$NH_{\pm}$};
\draw[color=qqqqff] (5.44,4.84) node {$dS_{\pm}$};
\draw[color=red] (3.26,2.24) node {S};
\draw[color=qqqqff] (3.26,6.76) node {C};
\draw[color=red] (7.46,2.24) node {$G_{\pm}$};
\draw[color=qqqqff] (7.46,6.76) node {$P_{\pm}$};
{\color{green}\draw[ ->,line width=1.3pt] (5.3,1) -- (1.6,1);}
{\color{green}\draw[ ->,line width=1.3pt] (7.1,2.24) -- (3.5,2.24);}
{\color{green}\draw[ ->,line width=1.3pt] (5.3,4.84) -- (1.6,4.84);}
{\color{green}\draw[ ->,line width=1.3pt] (7.1,6.84) -- (3.5,6.84);}
{\color{red}\draw[ ->,line width=1.3pt] (1.5,4.6) -- (1.5,1.1);}
{\color{red}\draw[ ->,line width=1.3pt] (5.3,4.6) -- (5.3,1.1);}
{\color{red}\draw[ ->,line width=1.3pt] (3.3,6.5) -- (3.3,2.5);}
{\color{red}\draw[ ->,line width=1.3pt] (7.46,6.5) -- (7.46,2.5);}
{\color{blue}\draw[ <-,line width=1.3pt] (3.1,2.2) -- (1.6,1);}
{\color{blue}\draw[ <-,line width=1.3pt] (7.1,2.2) -- (5.6,1);}
{\color{blue}\draw[ <-,line width=1.3pt] (3.1,6.7) -- (1.6,4.9);}
{\color{blue}\draw[ <-,line width=1.3pt] (7.2,6.7) -- (5.6,4.9);}
\end{tikzpicture}
\end{center}
\caption{Contractions in terms of {\color{green} compliance}, {\color{red}energy} and {\color{blue}mass}}
\end{figure}
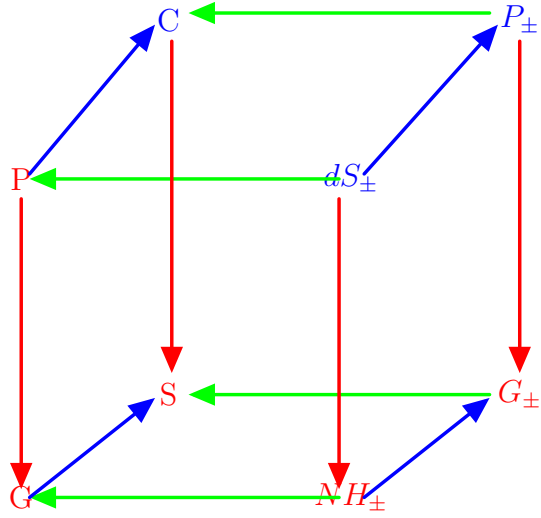
\begin{table}[htbp]
\begin{tabular}{|c|c|c|}
\hline
{\it Kinematical Process}&{\it Dynamical Process}&{\it Kinematical Lie Algebras}\\
\hline
Relative time groups&finite energy groups&$dS_{\pm}, P, P_{\pm}, C$\\
Absolute time groups&infinite energy groups&$NH_{\pm}, G, G_{\pm}, S$\\
Relative space groups&finite mass groups&$dS_{\pm}, NH_{\pm},P,G$\\
Absolute space&infinite mass groups&$P_{\pm}, G_{\pm}, C, S$\\
Cosmological groups&finite compliance groups&$dS_{\pm},NH_{\pm}, P_{\pm}, G_{\pm}$\\
Local groups&infinite compliance groups&$P, G,C,S$\\
\hline
\end{tabular}
\caption{Distribution of kinematical Lie algebras through the kinematical versus dynamical contractions}
\end{table}
\pagebreak
 \begin{table}[htbp]
\begin{tabular}{|c|c|c|c|c|c|c|c|}
\hline
Lie symbol&Lie algebra Name& $\{E,q_i\}$&$\{q_i,q_j\}$&$\{\pi_i,q_j\}$&$\{\pi_i,\pi_j\}$& $\{E,\pi_i\}$\\
\hline
 $dS_{\pm}$& $de~Sitter$ &$\frac{1}{m}\pi_i$ & $\frac{1}{mE_0}j_k\epsilon^k_{ij}$ & $\frac{E}{E_0}\delta_{ij}$ & $\mp \frac{1}{CE_0}j_k\epsilon^k_{ij}$& $\pm \frac{1}{C} q_i$\\
  $NH_{\pm}$& $Newton-Hooke$ & $\frac{1}{m}\pi_i$ & $0$ & $0$ & $0$& $\pm \frac{1}{C} q_i$ \\
 \hline
  $P$& $Poincare$ & $\frac{1}{m}\pi_i$ & $\frac{1}{mE_0}j_k\epsilon^k_{ij}$ & $\frac{E}{E_0}\delta_{ij}$ & $0$& $0$ \\
  $G$& $Galilei$ & $\frac{1}{m}\pi_i$ & $0$ & $0$ & $0$& $0$ \\
  \hline
  $P_{\pm}$& $Para~Poincare$ & $0$ & $0$ & $\frac{E}{E_0}\delta_{ij}$ & $\mp \frac{1}{CE_0}j_k\epsilon^k_{ij}$& $\pm \frac{1}{C} q_i$ \\
  $G_{\pm}$& $Para-Galilei$ & $0$ & $0$ & $0$ & $0$& $\pm \frac{1}{C} q_i$ \\
\hline
$C$& $Carroll$ & $0$ & $$ & $\frac{E}{E_0}\delta_{ij}$ & $0$& $0$ \\
 $S$& $Static$ & $0$ & $0$ & $0$ & $0$& $0$ \\
 \hline
\end{tabular}
\caption{The kinematical Poisson-Lie algebras}
\end{table}
\pagebreak
\begin{table}[htbp]
\center{}
\begin{tabular}{|c|c|c|c|}
\hline
~&$1^{st}$ order DE&$2^{nd}$ order DE&Lie algebras\\
\hline
$6$&$\frac{d\alpha_{\mu}}{d\varphi}=\alpha_{\nu}\epsilon^{\nu}_{\mu};~\alpha_{\mu}=j_{\mu},q_{\mu},\pi_{\mu}
$&$\frac{d^2\alpha_{\mu}}{d\varphi^2}=-\alpha_{\mu}$&All\\
\hline
$6$&$\frac{d}{dt}(q_i,\pi_i)=(\frac{\pi_i}{m},\pm\frac{q_i}{C})$&$\frac{d^2f}{dt^2}=\pm \frac{f}{mC}$&$dS_{\pm},NH_{\pm}$\\
$3$&$\frac{dq_i}{dt}=\frac{\pi_i}{m}$, ~~(cst~ velocity)&No~one&P, G\\
$3$&$\frac{d\pi_i}{dt}=\pm\frac{q_i}{C}$,~~( cst~ force)&No~one&$P_{\pm}$ ,$G_{\pm}$\\
\hline
$6$&$\frac{d}{dz}(j_{\mu},\pi_{\mu})=(\pi_{\nu},\mp\frac{j_{\nu}}{CE_0})\epsilon^{\nu}_{\mu},\frac{d}{dz}(q,E)=(\frac{E}{E_0},
\mp\frac{q}{C})$&$\frac{d^2f}{dz^2}=\mp\frac{f}{CE_0} $&$dS_{\pm}$ ,$P_{\pm}$\\
$3$&$\frac{dj_{\mu}}{dz}=\pi_{\nu}\epsilon^{\nu}_{\mu}$~(cst~pseudo-momentum),$\frac{dE}{dz}=\mp\frac{q}{C}$~(cst~force)&No~one&$NH_{\pm}$ ,$G_{\pm}$\\
$3$&$\frac{dj_{\mu}}{dz}=\pi_{\nu}\epsilon^{\nu}_{\mu}$~(cst~pseudo-momentum),$\frac{dq}{dz}=\frac{E}{E_0}$~(cst~number)&No~one&$P,C$\\
$2$&$\frac{dj_{\mu}}{dz}=\pi_{\nu}\epsilon^{\nu}_{\mu}$~(cst~pseudo-momentum)&No~one&G, S\\
\hline
$6$&$\frac{d}{dp}(j_{\mu},q_{\mu})=(q_{\nu},\frac{j_{\nu}}{mE_0})\epsilon^{\nu}_{\mu},\frac{d}{dp}(\pi,E)=-(\frac{E}{E_0},
\frac{\pi}{m})$&$\frac{d^2f}{dp^2}=-\frac{f}{mE_0} $&$dS_{\pm}$ ,P\\
$3$&$\frac{dj_{\mu}}{dp}=q_{\nu}\epsilon^{\nu}_{\mu}$~(cst~pseudo-position),$\frac{dE}{dp}=-\frac{\pi}{m}$~(cst~velocity)&No~one&$NH_{\pm}$, G\\
$3$&$\frac{dj_{\mu}}{dp}=q_{\nu}\epsilon^{\nu}_{\mu}$~(pseudo-position),$\frac{d\pi}{dp}=-\frac{E}{E_0}$~(cst~number)&No~one&P, C\\
$2$&$\frac{dj_{\mu}}{dp}=q_{\nu}\epsilon^{\nu}_{\mu}$,~(pseudo-position)&No~one&$G_{\pm}$, S\\
\hline
\end{tabular}
\caption{Equations of change}
\end{table}

\newpage
\section{Appendix}
This appendix serves to show how non additive parameters such as Lorentz boost are obtained from additive ones such as momentum. We use the brackets of table II.
\subsection{From momentum to boost}
The  Poincar\'e Lie algebra in one space dimension is defined by the Lie brackets
\begin{eqnarray}
[Q,P]=\frac{1}{E_0}H,~[Q,H]=\frac{1}{m}P,~[P,H]=0
 \end{eqnarray}
 where $Q$ genrates momenta, $P$ generates space translations while $P$ generates time translations. We verify that $exp(x_0^{\prime}P+t_0^{\prime}H)=Ad_{exp(pQ+xP+tH)}(exp(x_0P+t_0H))$ gives the Poincare space-time transformations
 \begin{eqnarray}\label{lorentz}
 x^{\prime}_0=\cosh(\frac{p}{\sqrt{mE_0}})x_0+\sqrt{\frac{E_0}{m}}\sinh(\frac{p}{\sqrt{mE_0}})t_0+x,
 ~t^{\prime}_0=\sqrt{\frac{m}{E_0}}\sinh(\frac{p}{\sqrt{mE_0}})x_0+\cosh(\frac{p}{\sqrt{mE_0}})t_0+t
 \end{eqnarray}
 where $p$ is an additive momentum.

  If we define the boost by $v=\sqrt{\frac{E_0}{m}}\tanh(\frac{p}{\sqrt{mE_0}})$, then we recover the corresponding non additive boosts composition law  $v^{\prime\prime}=\frac{v+v^{\prime}}{1+\frac{mvv^{\prime}}{E_0}}$. It is the usual Lorentz one when $E_0=mc^2$. Similary if  slowness is defined by $s=\sqrt{\frac{m}{E_0}}\tanh(\frac{p}{\sqrt{mE_0}})$, then the slowness composition law is $s^{\prime\prime}=\frac{s+s^{\prime}}{1+\frac{E_0vv^{\prime}}{m}}$.

   The limits of (\ref{lorentz}) are given in the table
 \begin{eqnarray}
 \begin{tabular}{|c|c|}
 \hline
 $E_0\rightarrow \infty$&$m \rightarrow \infty$\\
 \hline
 $ x^{\prime}_0=x_0+\frac{p}{m}t_0+x,~t^{\prime}_0=t_0+t$&$x^{\prime}_0=x_0+x,~ t^{\prime}_0=t_0+\frac{p}{E_0}x_0+t$\\
 \hline
 Galilei transformations&Carroll transformations\\
  on space-time&on space-time\\
 \hline
 \end{tabular}
 \end{eqnarray}
where $v=\frac{p}{m}$ is a Galilean boost and $s=\frac{p}{E_0}$ is a Carrollian slowness.
\subsection{From space translation for force}
The Lie algebra of one spatial Para-Poincar\'e Lie algebra $P_{\pm}$ is defined by the Lie brackets
 \begin{eqnarray}
[Q,P]=\frac{1}{E_0}H,~[[Q,H]=0,~[P,H]=\pm \frac{1}{C}Q.
 \end{eqnarray}
 We verify that $exp(p_0^{\prime}Q+t_0^{\prime}H)=Ad_{exp(pQ+xP+tH)}(exp(p_0Q+t_0H))$ gives the Para-Poincare momentum-time transformations
 \begin{eqnarray}\label{paralorentzminus}
 p^{\prime}_0=\cosh(\frac{x}{\sqrt{CE_0}})p_0-\sqrt{\frac{E_0}{C}}\sinh(\frac{x}{\sqrt{CE_0}})t_0+p,
 ~t^{\prime}_0=-\sqrt{\frac{C}{E_0}}\sinh(\frac{x}{\sqrt{CE_0}})p_0+
 \cosh(\frac{x}{\sqrt{CE_0}})t_0+t
 \end{eqnarray}
 in the $P_-$ case and
 \begin{eqnarray}\label{paralorentzplus}
 p^{\prime}_0=\cos(\frac{x}{\sqrt{CE_0}})p_0+\sqrt{\frac{E_0}{C}}\sin(\frac{x}{\sqrt{CE_0}})t_0+p,
 ~t^{\prime}_0=-\sqrt{\frac{C}{E_0}}\sin(\frac{x}{\sqrt{CE_0}})p_0+\cos(\frac{x}{\sqrt{CE_0}})t_0+t
 \end{eqnarray}
 in the $P_+$ case. The additive parameter $x$ is non compact (compact) in the $P_-(P_+)$ case.\\
 If  $f=\sqrt{\frac{E_0}{C}}\tanh(\frac{x}{\sqrt{CE_0}})$ ($f=\sqrt{\frac{E_0}{C}}\tan(\frac{x}{\sqrt{CE_0}})$) is a force for $P_-$ ($P_+$) while $\phi=\sqrt{\frac{C}{E_0}}\tanh(\frac{x}{\sqrt{CE_0}})$ ($\phi=\sqrt{\frac{C}{E_0}}\tan(\frac{x}{\sqrt{CE_0}})$) is an inverse of force for $P_-$ ($P_+$), then we get the non additive composition laws $f^{\prime\prime}=\frac{f+f^{\prime}}{1\mp\frac{Cff^{\prime}}{E_0}}$ and $\phi^{\prime\prime}=\frac{\phi+\phi^{\prime}}{1\mp\frac{E_0\phi \phi^{\prime}}{C}}$ for the $P_{\pm}$ case. Moreover we have that
 \begin{eqnarray}
 \begin{tabular}{|c|c|}
 \hline
 $E_0\rightarrow \infty$&$C \rightarrow \infty$\\
 \hline
 $ p^{\prime}_0=p_0 \pm\frac{x}{C}t_0+p,~t^{\prime}_0=t_0+t$&$p^{\prime}_0=p_0+p,~ t^{\prime}_0=t_0-\frac{x}{E_0}p_0+t$\\
 \hline
 Para-Galilei transformations&Carroll transformations\\
  on momentum-time&on momentum-time\\
 \hline
 \end{tabular}
 \end{eqnarray}
 where $f=\pm\frac{x}{C}$ is a force for the Para-Galilei case $P_{\pm}$ an force and $\phi=-\frac{x}{E_0}$ is a Carrollian inverse of force.
 \subsection{From time translation to dampinglike coefficient}
The Lie algebra of one spatial Newton-Hooke Lie algebra $NH_{\pm}$ is defined by the Lie brackets
 \begin{eqnarray}
[Q,P]=0,~[Q,H] =\frac{1}{m}P,~[P,H]=\pm \frac{1}{C}Q
 \end{eqnarray}
 We verify that $exp(p_0^{\prime}Q+x_0^{\prime}P)=Ad_{exp(pQ+xP+tH)}(exp(p_0Q+x_0P))$ gives the Newton-Hooke momentum-space transformations
 \begin{eqnarray}\label{newtonhookeplus}
 p^{\prime}_0=\cosh(\frac{t}{\sqrt{mC}})p_0-\sqrt{\frac{m}{C}}\sinh(\frac{t}{\sqrt{mC}})x_0+p,~
 x^{\prime}_0=-\sqrt{\frac{C}{m}}\sinh(\frac{t}{\sqrt{mC}})p_0+\cosh(\frac{t}{\sqrt{mC}})x_0+x
 \end{eqnarray}
 in the $NH_+$ case and
 \begin{eqnarray}\label{newtonhookeminus}
p^{\prime}_0=\cos(\frac{t}{\sqrt{mC}})p_0+\sqrt{\frac{m}{C}}\sin(\frac{t}{\sqrt{mC}})x_0+p,
~x^{\prime}_0=-\sqrt{\frac{C}{m}}\sin(\frac{t}{\sqrt{mC}})p_0+\cos(\frac{t}{\sqrt{mC}})x_0+x
 \end{eqnarray}
 in the $NH_-$ case. The additive parameter $t$ is non compact (compact) in the $NH_+ (NH_-)$ case. Moreover the dampinglike coefficient  $b=\sqrt{\frac{m}{C}}\tanh(\frac{t}{\sqrt{mC}})$ ($b=\sqrt{\frac{m}{C}}\tan(\frac{t}{\sqrt{mC}})$) for the $NH_+$($NH_-$) whose the dimension is $MT^{-1}$ and  $\beta=\sqrt{\frac{C}{m}}\tanh(\frac{t}{\sqrt{mC}})$ and an inverse of a dampinglike ($\beta=\sqrt{\frac{C}{m}}\tan(\frac{t}{\sqrt{mC}})$) for the $NH_+$($NH_-$) whose the dimension is $M^{-1}T$ satisfy the non additive composition laws $b^{\prime\prime}=\frac{b+b^{\prime}}{1\pm\frac{Cbb^{\prime}}{m}}$ and $\beta^{\prime\prime}=\frac{\beta+\beta^{\prime}}{1\pm\frac{m\beta\beta^{\prime}}{C}}$.\\
 for the $NH_{\pm}$ cases.\\
 We finally have
 \begin{eqnarray}
 \begin{tabular}{|c|c|}
 \hline
 $m\rightarrow \infty$&$C \rightarrow \infty$\\
 \hline
 $ p^{\prime}_0=p_0 \mp \frac{t}{C}x_0+p,~x^{\prime}_0=x_0+x$&$p^{\prime}_0=p_0+p,~ x^{\prime}_0=x_0-\frac{t}{m}p_0+x$\\
 \hline
 Para-Galilei group action&Galilei group action\\
 on momentum-space&on momentum-space\\
 \hline
 \end{tabular}
 \end{eqnarray}

\newpage
\begin{thebibliography}{99}
\bibitem{jmll}H.Bacry, J.M.L\'evy-Leblond,
``Possible kinematics,'' {\it J.Math.Phys., $\textbf{9}(1968),1605-1614$}.doi.org/10.1063/1.1664490.
\bibitem{inonuwigner}E.Inonu and E.P.Wigner , On the contractions of groups and their representations, {\it Proc. Nat.Acad.Sc., $\textbf{39}(1953), 510-524$}.doi.org/10.1073/pnas.39.6.510
\bibitem{saletan}E.Saletan, Contraction of Lie groups, {\it J.Math.Phys., $\textbf{2}(1961),1$}.doi.org/10.1063/1.1724208.
\bibitem{jmll2}J.M.L\'evy-Leblond, Une nouvelle limite non relativiste du groupe de Poincar\'e, {\it AIHPA $\textbf{3(1)}(1965), 1-12$}.
 \bibitem{sanjuan} M.A.F. Sanjuan, Group contraction and the nine Cayley-Klein geometries, {\it Int.J.Theor.Phys. $\textbf{23(1)}(1984),1-14$}.doi.org/10.1007/BF02080668.
 \bibitem{dooley}Dooley, A. H. and J. W. Rice, On contractions of semisimple Lie groups, {\it Transactions of the American Mathematical Society $\textbf{289(1)}(1985),  185-202$}.doi.org/10.1090/S0002-9947-1985-0779059-4.
 \bibitem{gromov}N.A.Gromov, Transitions: Contractions and analytical continuations of the Cayley-Klein Groups, {\it Int.J.Theor.Phys. $\textbf{29(6)}(1990),607-620$}.doi.org/10.1007/BF006772035.
\bibitem{fialowski}A.Fialowski and M.de Montigny, Contraction and Deformation of Lie algebras, {\it J.Phys.A: Math.Gen. $\textbf{38}(2005),6335-6349$}.doi.org/10.1088/0305-4470/38/28/006.
 \bibitem{mcrae} A.S.McRae , Clifford algebras and possible kinematics, {\it SIGMA $\textbf{3}(2007),79$}.doi.org/10.3842/SIGMA.2007.079[http://arxiv.org/abs/0707.2869]
 \bibitem{guo}H.Y.Guo, C.G. Huang, H.T. Wu and B. Zhou, The principle of relativity, kinematics and algebraic relations, {\it Science China Physics, Mechanics and Astronomy $\textbf{53(4)}(2010),591-597$}.DOI : 10.1007/s11433-010-0162-6[arXiv:0812.0871[hep-th]]
\bibitem{ancilla1} A.Ngendakumana,J.Nzotungicimpaye and L.Todjhounde, Group theoretical construction of planar noncommutative phase spaces, {\it J.Math.Phys. $\textbf{55}(2014),013508-$}.doi.org/10.1063/1.4862843[https://arxiv:1308.3065v1[math-phys]]
\bibitem{campaomor2010}R.Campaomor-Stursberg and M.R.de Traubeberg, Contraction-based classification of supersymmetric extension of kinematical Lie algebras, {\it Phys. Atom. Nucl. $\textbf{73}(2010) , 264-268$}.doi.org/10.1134/S1063778810020109.
  \bibitem{rembielinski} J.Rembielinski and W Tybor, Possible superkinematics, {\it Acta Phys. Polon., B$\textbf{\textbf{15}}()1984$ ,611-615}.
  \bibitem{chao}H.Chao-Guang and Li Lin , Possible supersymmetric kinematics, {\it Chin. Phys. C $\textbf{39}(9)(2015)093103$}.doi.org/10.1088/1674-1137/39/9/093103.[ArXiv:1409.5498[hep-th]].
\bibitem{barducci1}A.Barducci, R.Casalbuoni and J.Gomis, Vector SUSY models with Carroll or Galilei invariance, {\it Phys.Rev.D $\textbf{99}(2019) , 045016$ }.doi.org/10.1103/PhysRevD.99.045016. [ArXiv:1811.12672].
\bibitem{derome}J.R.Derome and J.G.Dubois, Hooke's Symmetries and Nonrelativistic Cosmolgical Kinematics,{ \it Il Nuovo Cimento B, $\textbf{9(2)}(1972),351-376$}.doi.org/10.1007/BF02734453.
\bibitem{ancilla} A.Ngendakumana, J.Nzotungicimpaye and L.Todjhounde, Para-Galilean versus Galilean Noncommutative Spaces, {\it Int.J.Geom.Meth. Mod.Phys. $\textbf{10(10)}(2013),1350049$.}doi.org/10.1142/S0219887813500497[arXiv:1207.3919[math-phys]].
  \bibitem{gibbonsthoughts} G.~W.~Gibbons,Thoughts on tachyon cosmology, {\it
  Class. Quant. Grav.} $\textbf{20}(2003),S321-S346$.https://doi.org/10.1088/0264-9381/20/12/301.[hep-th/0301117].
   \bibitem{gibbonscondensates}G.~W.~Gibbons,K.~Hashimoto and P.Yi, Tachyon Condensates, Carrollian Contraction of Lorentz Group, and Fundamental Strings, {\it JHEP}(2002)061. doi:10.1088/1126-6708/2002/09/062.
     \bibitem{escamilla}C.Escamilla-Rivera, G.Garcia-Jimenez and O.Obregon, Unveilling the tachyons dynamics in the Carrollian limit, {\it Rev.Mex.Fis.E $\textbf{56(2)}(2010),177-180$}.[arXiv:1009.0105].
 \bibitem{duval}C.Duval, G.W.Gibbons, P.A.Horvathy and P.M.Zhang, Carroll versus Newton and Galilei:two dual non-Einsteinian concepst of time,{\it Class.Quantum.Grav, $\textbf{31}(2014), 085016-$}.doi.org/10.1088/0264-9381/31/8/085016 [arxiv:1402.065[gr-qc]].
 \bibitem{Lebellac}M.Le Bellac and J.M.L\'evy-Leblond , Galilean electromagnetism, {\it Nuovo Cimento $\textbf{14B}(1973), 217-234$.}doi.org/10.1007/BF02895715.
  \bibitem{joakimgomis}E. Bergschoeff ,J.Gomis and G.Longhi, Dynamics of Carroll particles,{\it Classical Quantum Gravity} $\textbf{31}$(2014),205009.doi.org/10.1088/0264-9381/31/20/205009.[arxiv:1405.2264[hep-th]].
  \bibitem{cardona}B.Cardona, J.Gomis and J.M.Pons, Dynamics of Carroll srings, {\it JHEP $textbf{7}$}(2016)50.doi.org/10.1007/JHEP07(2016)050.[ArXiv:1605.05483].
  \bibitem{Sou73}J-M. Souriau,Le milieu \'elastique soumis aux ondes gravitationnelles, \textit{Ondes et radiations gravitationnelles},
Colloques Internationaux du CNRS No 220, p. 243. Paris (1973).
\bibitem{Carroll4GW} C.~Duval, G.~W.~Gibbons, P.~A.~Horvathy and P.-M.~Zhang, Carroll symmetry of plane gravitational waves,Class. Quant. Grav. $\textbf{34} (2017)$.doi.org/10.1088/1361-6382/aa7f62. [arXiv:1702.08284 [gr-qc]].
\bibitem{hartong}J.Hartong, Gauging the Carroll algebra and ultra-relativistic gravity. JHEP \textbf{2015(8)}69.doi.org/10.1007/JHEP08(2015)069.[arxiv.1505.05011[hep-th]].
\bibitem{bekaert1}X.Bekaert and K.Morand (2018), Connections and dynamical trajectories in generalised Newton-Cartan gravity I. An intrinsic view perspective {\it J.Math.Phys.}\textbf{57(2)}(2016)022507.doi.org/10.1063/1.4937445.
\bibitem{bekaert2}X.Bekaert and K.Morand (2018), Connections and dynamical trajectories in generalised Newton-Cartan gravity II. An ambient perspective {\it J.Math.Phys.}\textbf{59(7)}(2018)072503.doi.org/10.1063/1.5030328.
  \bibitem{morand} K.Morand, Embedding Galilean and Carrolian geometries. gravitational waves, {\it Phys. Review D $\textbf{88(6)}(2018), 063008$}.[arXiv:1811.12681[hep-th]].
\bibitem{DGH91}C. Duval, G.W. Gibbons, P. Horvathy, Celestial mechanics, conformal structures and gravitational waves,
{ \it Phys.Rev} {\textbf D43} (1991).3907-3922.doi.org/10.1103/PhysRevD.43.3907. [hep-th/0512188].
\bibitem{barducci2}A.Barducci, R.Casalbuoni and J.Gomis, Confined dynamical systems with Carroll and Galilean symmetries, {\it Phys.Rev.D $\textbf{98}(2018) , 085018$ }.doi.org/10.1103/PhysRevD.98.085018. [ArXiv:1804.10495[hep-th]].
 \bibitem{bergshoef}E.Bergshoef,J.Gomis,B.Rollier and J.Rosseel, Carroll versus Galilei gravity, {\it JHEP, \textbf{03}(2017), 165}. doi.org/10.1007/JHEP03(2017)165.[arXiv:1402.0657[gr-qc]].
\bibitem{ciambelli}L.Ciambelli, C.Marteau, A.C.Petkou, P.M.Petropoulos and K.Siampos, Covariant Galilean versus Carrollian hydrodynamics from relativistic fluids,{\it Class. Quantum Grav.} 35(2018)165001. doi.org/10.1088/1361-6382/aacf1a. [ArXiv:1802.05286[hep-th]].
\bibitem{dyson} F.J.Dyson ,  Missed opportunities,{\it Bulletin of the American Mathematical Society, $\textbf{78(5)}(1972),635-652$}. doi.org/10.1090/S0002-9904-1972-12971-9.


.




\end{thebibliography}
\end{document}